\documentclass[useAMS,usenatbib]{mn2e}
\usepackage{psfig}
\usepackage{multirow}
\usepackage{subfigure}

\providecommand{\adsurl}[1]{\href{#1}{ADS}}

\newcommand{\eg}{{\it e.g.~}}
\newcommand{\ie}{{\it i.e.~}}

\newcommand{\be}{\begin{equation}}
\newcommand{\ee}{\end{equation}}
\newcommand{\ba}{\begin{eqnarray}}
\newcommand{\ea}{\end{eqnarray}}
\newcommand{\brr}{\begin{array}}

\newcommand{\err}{\end{array}}
\newcommand{\bc}{\begin{center}}
\newcommand{\ec}{\end{center}}

\def\mnras{MNRAS}

\def\grtsim{\mathrel{\hbox{\rlap{\hbox{\lower2pt\hbox{$\sim$}}}\raise2pt\hbox{$>$}}}} \def\lesssim{\mathrel{\hbox{\rlap{\hbox{\lower2pt\hbox{$\sim$}}}\raise2pt\hbox{$<$}}}}

\title[Weak lensing in coupled dark energy cosmologies]{Weak lensing predictions for coupled dark energy cosmologies at non-linear scales}
\author[E. Beynon, M. Baldi, D.J. Bacon, K. Koyama, C. Sabiu]{\parbox[t]{\textwidth}{Emma Beynon$^{1}$\thanks{emma.beynon@port.ac.uk}, Marco Baldi$^{2,3}$\thanks{marco.baldi@universe-cluster.de}, David J. Bacon$^{1}$, Kazuya Koyama$^{1}$, Cristiano Sabiu$^{4}$}\\
\vspace*{4pt} \\
$^{1}$ Institute of Cosmology and Gravitation, University of Portsmouth, Dennis Sciama Building, Portsmouth, PO1 3FX, UK \\
$^{2}$ Excellence Cluster Universe, Boltzmannstr. 2, D-85748 Garching, Germany\\
$^{3}$ University Observatory, Ludwig-Maximillians University Munich, Scheinerstr. 1, D-81679 Munich, Germany\\
$^{4}$  Department of Physics \& Astronomy, University College London, Gower Street, London, United Kingdom\\}

\begin{document}

\date{Accepted ---. Received ---; in original form ---}

\pagerange{\pageref{firstpage}--\pageref{lastpage}} \pubyear{2010}

\maketitle

\label{firstpage}

\begin{abstract}
We present non-linear weak lensing predictions for coupled dark energy models using the {\small CoDECS} simulations. We calculate the shear correlation function and error covariance expected for these models, for forthcoming ground-based (such as DES) and space-based (Euclid) weak lensing surveys. We obtain predictions for the discriminatory power of a ground-based survey similar to DES and a space-based survey such as Euclid in distinguishing between $\Lambda$CDM and coupled dark energy models; we show that using the non-linear lensing signal we could discriminate between $\Lambda$CDM  and exponential constant coupling models with $\beta_0\geq0.1$ at $4\sigma$ confidence level with a DES-like survey, and $\beta_0\geq0.05$ at $5\sigma$ confidence level with Euclid. We also demonstrate that estimating the coupled dark energy models' non-linear power spectrum, using the $\Lambda$CDM Halofit fitting formula, results in biases in the shear correlation function that exceed the survey errors.
\end{abstract}

\begin{keywords}
Gravitation; Gravitational Lensing; Cosmology: Theory
\end{keywords}

\section{Introduction}\label{intro}

After more than a decade of continuous improvements in the accuracy of cosmological observations  -- which has led to the establishment of a broadly accepted representation of our Universe known as the {\em Concordance Cosmological Model} (CCM) -- we are now entering the epoch of precision cosmology. The great wealth of high-precision cosmological data expected throughout the next few years offers the exciting prospect of tightly constraining the parameters of the CCM or possibly detecting deviations from its predictions.

The unprecedented resolution of the Planck satellite \citep[][]{Ade:2011ah} in measuring the angular temperature fluctuations and the polarization of the Cosmic Microwave Background
 (CMB) radiation will extract a significant amount of new information beyond that yielded by previous CMB experiments, affording very tight constraints on the initial conditions of the Universe.
 
  At the other end of the cosmic expansion history, \ie at low redshifts, present and future surveys measuring the clustering of sources \citep[][]{Euclid-r} and gravitational lensing effects \citep[][]{Massey:2007gh,Euclid-r} will greatly improve our  knowledge of the spatial distribution of baryonic and cold dark matter (CDM) in the local Universe, and of its time evolution, providing new tests of the gravitational instability processes driving the growth of cosmic structures.

 In particular, one of the most mysterious phenomena characterizing the low-redshift Universe is the appearance of a Dark Energy (DE) component capable of driving the observed present acceleration of the cosmic expansion \citep[][]{Riess_etal_1998,Perlmutter_etal_1999,Schmidt_etal_1998,snls,kowalski_etal_2008}, and required to also explain a number of other observations, \eg the angular power spectrum of CMB temperature anisotropies \citep[][]{wmap5,wmap7}, the evolution of the number counts of massive galaxy clusters as a function of redshift \citep[][]{Borgani_2006,Vikhlinin_etal_2009b,Mantz_etal_2010}, the angular correlation of galaxies in large galaxy surveys \citep[][]{Percival_etal_2001,Cole_etal_2005,Reid_etal_2010}, or the observed scale of the Baryon Acoustic Oscillations (BAO) \citep[][]{Percival_etal_2009}. In the standard $\Lambda $CDM model, such a DE component is identified with a cosmological constant $\Lambda $, a quantity with negative pressure and constant energy density throughout the whole expansion history of the Universe, and with no spatial fluctuations.

 This simple picture is very successful in reproducing a wide range of observational data. However, the cosmological constant scenario suffers from serious conceptual problems concerning the extremely small value of  the constant DE density as compared to
 the typical densities of the early Universe, known as the ``fine tuning problem" \citep[see e.g.][]{Weinberg_1988}, and the apparent coincidence that it dominates over CDM only at relatively recent cosmological epochs, the ``coincidence problem" \citep[see e.g.][]{Huey_Wandelt_2006}. In order to overcome these problems, alternative models based on the dynamic evolution of a classical scalar field have been proposed \citep[][]{Wetterich_1988,Ratra_Peebles_1988,ArmendarizPicon_etal_2000}. Abandoning the simple picture of a cosmological constant, however, necessarily requires us to consider and to include in our models of the Universe the presence of spatial fluctuations and of possible interactions of the new physical degree of freedom represented by the DE scalar field.

 It is in this context that models of interacting DE have been proposed as a natural extension of the minimally coupled dynamic scalar field scenario \citep[][]{Wetterich_1995,Amendola_2000,Farrar2007}. Although an interaction of the DE scalar field with baryonic particles is tightly constrained by observations \citep[][]{Hagiwara_etal_2002},
 the same bounds do not apply to the case of a selective interaction between DE and CDM, as first speculated by \cite{Damour_Gibbons_Gundlach_1990}, which has therefore received substantial attention as a realistic competitor to the standard $\Lambda $CDM model.

 Various different forms of interactions between DE and CDM particles (including massive neutrinos) have been proposed and investigated in the literature \citep[as \eg by][]{Amendola_2004,CalderaCabral:2008bx,Pettorino_Baccigalupi_2008,Amendola_Baldi_Wetterich_2008,Boehmer:2009tk,Koyama_etal_2009,Honorez_etal_2010},
 and their impact on the linear growth of density perturbations \citep[see \eg][]{DiPorto_Amendola_2008,CalderaCabral:2009ja,Valiviita:2008iv,Majerotto:2009np,Valiviita:2009nu,Clemson:2011an} and on the nonlinear regime of structure formation \citep[][]{Maccio_etal_2004,Baldi_etal_2010,Baldi_2011a,Li_Barrow_2011,Baldi_Pettorino_2010,Li:2010zzx,Li:2010eu}
 has been extensively studied in recent years. For many such models, robust and realistic observational constraints on the interaction strength have been derived
 based on CMB and LSS data \citep[][]{Bean_etal_2008,LaVacca_etal_2009,xia_2009}, local dynamical tests \citep[][]{Kesden_Kamionkowski_2006,Keselman_Nusser_Peebles_2009}, and Lyman-$\alpha $ observables \citep[][]{Baldi_Viel_2010}.  Although these observational bounds have strongly restricted the allowed parameter space for interacting DE cosmologies, none of them has yet been able to rule out the model, or to unambiguously detect the presence of a DE-CDM interaction with compelling statistical significance.

 In this respect, exciting times are ahead of us, with the realistic possibility of exploiting the joint power of forthcoming high-precision cosmological observations to break many of the existing degeneracies between competing cosmological models and finally disentangle the distinctive features of alternative scenarios. Dark energy interactions will be one of the issues that can be tested, and so the next generation of cosmological data will possibly provide a real indication of the nature of the DE phenomenon.

In the present paper, we examine the usefulness of weak gravitational lensing for discriminating between interacting dark energy models. We wish to show how the lensing signal depends on the dark energy interaction, and whether this dependence is sufficiently strong that it could be detected with forthcoming lensing surveys. In particular, we will provide forecasts for the capability of future large Weak Lensing (WL) surveys --both a ground-based survey similar to the {\em Dark Energy Survey} (DES)\footnote[1]{http://www.darkenergysurvey.org},
 and a space-based survey, i.e. EUCLID\footnote[2]{http://www.ias.u-psud.fr/imEuclid} -- to detect a DE-CDM interaction. Our particular focus in this paper is the non-linear regime, as this regime provides much of the power for lensing. To this end, we exploit the full non-linear matter power spectrum as predicted by the
 {\small CoDECS} simulations \citep[][]{CoDECS}, the largest suite of self-consistent and high-resolution N-body simulations for interacting DE cosmologies to date. 

 The paper is organized as follows: in Sec.~\ref{coupledDE} we describe the main features of the interacting DE models under investigation; in Sec.~\ref{lenscoupled} we discuss gravitational lensing in the context of interacting DE models, and in Sec.~\ref{Simulations} we describe the methods used to compute the necessary nonlinear power spectra. In Sec.~\ref{Results} we present the results of our analysis, giving forecasts for forthcoming lensing surveys; we draw our conclusions in Sec.~\ref{Conclusions}.

\section{Coupled dark energy models}\label{coupledDE}

Coupled DE (cDE) models have been widely investigated in the literature concerning their cosmological background evolution
as well as the behaviour of linear and nonlinear density perturbations in these models  \citep[see \eg][and references therein]{Amendola_2000,Amendola_2004,Pettorino_Baccigalupi_2008,
DiPorto_Amendola_2008,Baldi_etal_2010,Li_Barrow_2011,Baldi_2011a}. Here we only briefly introduce the definitions and the notation
that will be assumed throughout the paper for the different cDE models; we refer the interested reader to the literature above for a thorough
discussion of cDE scenarios.

In the present work, we will consider cDE models defined by the following set of background dynamic equations:
\begin{eqnarray}
\label{klein_gordon}
\ddot{\phi } + 3H\dot{\phi } +\frac{dV}{d\phi } &=& \sqrt{\frac{2}{3}}\beta _{c}(\phi ) \frac{\rho _{c}}{M_{{\rm Pl}}} \,, \\
\label{continuity_cdm}
\dot{\rho }_{c} + 3H\rho _{c} &=& -\sqrt{\frac{2}{3}}\beta _{c}(\phi )\frac{\rho _{c}\dot{\phi }}{M_{{\rm Pl}}} \,, \\
\label{continuity_baryons}
\dot{\rho }_{b} + 3H\rho _{b} &=& 0 \,, \\
\label{continuity_radiation}
\dot{\rho }_{r} + 4H\rho _{r} &=& 0\,, \\
\label{friedmann}
3H^{2} &=& \frac{1}{M_{{\rm Pl}}^{2}}\left( \rho _{r} + \rho _{c} + \rho _{b} + \rho _{\phi} \right)\,,
\end{eqnarray}
where an overdot represents a derivative with respect to the cosmic time $t$, $H\equiv \dot{a}/a$ is the
Hubble function, $V(\phi )$ is the scalar field self-interaction potential, $M_{\rm Pl}\equiv 1/\sqrt{8\pi G}$
is the reduced Planck Mass, and the subscripts $b\,,c\,,r$, indicate baryons, CDM, and radiation, respectively.

The function $\beta _{c}(\phi )$ sets the direction and the strength of the energy-momentum flow between the DE scalar field $\phi $
and the CDM fluid, while the function $V(\phi )$ determines the dynamical evolution of the DE density. In the present work we will consider
two possible choices for each of these two functions, namely an exponential \citep[][]{Lucchin_Matarrese_1984,Wetterich_1988} and a SUGRA \citep[][]{Brax_Martin_1999}
potential,
\begin{eqnarray}
{\rm EXP:}&\quad & V(\phi ) = Ae^{-\alpha \phi } \,,\\
{\rm SUGRA:}&\quad & V(\phi ) = A\phi ^{-\alpha }e^{\phi ^{2}/2} \,,
\end{eqnarray}
where $\alpha $ is a positive constant and where for simplicity the field $\phi $ has been expressed in units of the reduced Planck mass $M_{\rm Pl}$,
as well as both a constant and an exponentially growing coupling function $\beta _{c}(\phi )$:
\begin{equation}
\beta _{c}(\phi ) = \beta _{0}e^{\beta _{1}\phi }\,,
\end{equation}
characterized by $\beta _{1}=0$ and $\beta _{1}>0$, respectively. The most relevant difference between the exponential potential and the SUGRA potential
relies on the fact that the latter features a global minimum at finite scalar field values; this allows for a change of direction of the scalar field motion,
which is the main feature of the recently proposed ``Bouncing cDE" scenario \citep[][]{Baldi_2011c}.
One should also notice that the notation introduced in Eqs.~(\ref{klein_gordon}-\ref{friedmann}) corresponds to the original convention proposed by \citet{Amendola_2000}
and has been adopted by several other studies, including the {\small CoDECS} project considered in the present work, but it differs by a constant
factor $\sqrt{2/3}$ from what is used in another part of the related literature \citep[as \eg][]{Pettorino_Baccigalupi_2008,Baldi_etal_2010}. The specific models
considered in the present work have been described in full detail by \citet{Baldi_2011c} and \citet{CoDECS}; we summarize them in Table~\ref{tab:models},
where the features and the specific parameters of each model are outlined.
\begin{table*}
\begin{tabular}{llccccc}
\hline
Model & Potential  &
$\alpha $ &
$\beta _{0}$ &
$\beta _{1}$ &
$w_{\phi }(z=0)$ &
$\sigma _{8}(z=0)$\\
\\
\hline
$\Lambda $CDM & $V(\phi ) = A$ & -- & -- & -- & $-1.0$ &  $0.809$ \\
EXP001 & $V(\phi ) = Ae^{-\alpha \phi }$  & 0.08 & 0.05 & 0 & $-0.997$ & $0.825$ \\
EXP002 & $V(\phi ) = Ae^{-\alpha \phi }$  & 0.08 & 0.1 & 0 & $-0.995$ & $0.875$ \\
EXP003 & $V(\phi ) = Ae^{-\alpha \phi }$  & 0.08 & 0.15 & 0 & $-0.992$ & $0.967$\\
EXP008e3 & $V(\phi ) = Ae^{-\alpha \phi }$  & 0.08 & 0.4 & 3 & $-0.982$ & $0.895$ \\
SUGRA003 & $V(\phi ) = A\phi ^{-\alpha }e^{\phi ^{2}/2}$  & 2.15 & -0.15 & 0 & $-0.901$ & $0.806$ \\
\hline
\end{tabular}
\caption{Interacting dark energy models considered in this work. In addition to the concordance $\Lambda$CDM model, we consider the exponential potential with three interaction strengths; the exponential potential with a time-varying strength; and the SUGRA potential.}
\label{tab:models}
\end{table*}
\ \\

The evolution equations for linear density perturbations in the context of a cDE cosmology have been derived in the literature
\citep[see \eg][]{Amendola_2004,Pettorino_Baccigalupi_2008}, and in the Newtonian limit of sub-horizon scales can be expressed
as follows:
\begin{eqnarray}
\label{gf_c}
\ddot{\delta }_{c} &=& -2H\left[ 1 - \beta _{c}\frac{\dot{\phi }}{H\sqrt{6}}\right] \dot{\delta }_{c} + 4\pi G \left[ \rho _{b}\delta _{b} + \rho _{c}\delta _{c}\Gamma _{c}\right] \,, \\
\label{gf_b}
\ddot{\delta }_{b} &=& - 2H \dot{\delta }_{b} + 4\pi G \left[ \rho _{b}\delta _{b} + \rho _{c}\delta _{c}\right]\,,
\end{eqnarray}
where $\delta _{c,b}\equiv \delta \rho _{c,b,}/\rho _{c,b}$ are the relative density perturbations of the coupled CDM and uncoupled baryonic fluids, respectively,
and where the scalar field dependence of the coupling function $\beta _{c}(\phi )$ has been omitted for simplicity.
In Eq.~(\ref{gf_c}), the factor $\Gamma _{c}\equiv 1 + 4\beta _{c}^{2}(\phi )/3$ represents an
additional fifth-force mediated by the DE scalar field $\phi $ for CDM perturbations, while the
second term in the first square bracket at the right-hand-side of Eq.~(\ref{gf_c}) is an extra friction term
on CDM fluctuations arising as a consequence of momentum conservation \citep[see e.g.][for a
derivation of Eqs.~(\ref{klein_gordon}-\ref{friedmann},\ref{gf_c},\ref{gf_b}) and for a
detailed discussion of the extra friction and fifth force corrections to the evolution of linear perturbations]{Amendola_2004,
Pettorino_Baccigalupi_2008,Baldi_etal_2010,Baldi_2011b}.
As a consequence of these two additional terms in the perturbed dynamic equations, CDM fluctuations will grow faster in cDE models
with respect to a standard $\Lambda $CDM cosmology, thereby reaching a higher $\sigma _{8}$ normalization at $z=0$ if starting from the same
amplitude at the last scattering surface $z_{\rm CMB}\approx 1100$, as shown in the last column of Table~\ref{tab:models}. However, in the nonlinear regime
the interplay between the friction term and the fifth force is not so straightforward as for the case of linear perturbations, due to the fact that
as a consequence of virialization processes, the local velocity field will not necessarily be aligned to the local gradient of the
gravitational potential, as one can see from the three-dimentional generalization of Eq.~(\ref{gf_c}) to a system of point-like massive particles:
\begin{equation}
\dot{\vec{v}}_{c} = \beta _{c}(\phi )\frac{\dot{\phi }}{\sqrt{6}}\vec{v}_{c} - \vec{\nabla }\left[ \sum_{c}\frac{GM_{c}(\phi )\Gamma _{c}}{r_{c}} + \sum_{b}\frac{GM_{b}}{r_{b}}\right] \,,
\end{equation}
where $r_{c,b}$ is the physical distance of the target coupled particle from the other CDM and baryonic particles, respectively.
The effect of the friction term in the nonlinear regime has been shown to induce a suppression of small-scale power in the cDE models
with respect to the nonlinear power that would be inferred based on the large-scale $\sigma _{8}$ normalization in the context of a $\Lambda $CDM universe
\citep[][]{Baldi_2011b,CoDECS}.
Such suppression will have important consequences on the weak lensing constraints on cDE models that we want to address in the present work. Therefore, although it is possible to estimate the full matter power in cDE scenarios by applying nonlinear corrections (calibrated on $\Lambda $CDM simulations) to the re-normalized linear power spectrum \citep[as recently done e.g. by][]{Amendola_etal_2011}, in order to reach high accuracy at scales relevant for present and future large lensing surveys it is necessary to rely on a fully nonlinear treatment of cDE scenarios via specific N-body simulations. A discussion on the comparison between these two approaches is presented in Section \ref{comparison}.

\begin{figure}
\hspace{7mm}
\psfig{figure=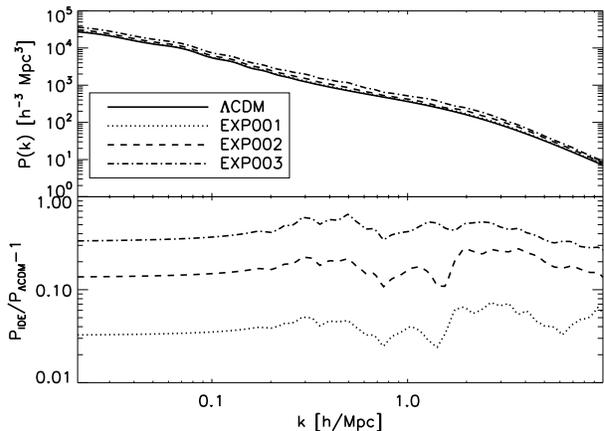,width=76mm}
 \caption{Power spectrum for $\Lambda$CDM and cDE models with constant coupling at z=0.}
 \label{fig:PowerspectrumConstant}
\end{figure}

Figure \ref{fig:PowerspectrumConstant} shows the power spectra for each of the constant coupling ($\beta_1=0$) models normalised by WMAP7. The values of these couplings were chosen since cDE models with $\beta_0 \leq 0.15$ can fit the angular diameter distance to decoupling measured by WMAP7, so these are of particular interest as they are consistent with current observations of the background, but may on the other hand affect the growth of structures. It can be seen that there is a 2-7\% difference in the $z=0$ power spectrum between $\Lambda$CDM and EXP001, the lowest of the couplings investigated here, and a 25-65\% difference between $\Lambda $CDM and the highest of the couplings, EXP003. 

In this analysis we do not use the simulated power spectrum directly but instead use the ratio between the $\Lambda$CDM and cDE power spectra to find the difference in the growth of modes for different couplings with the same initial conditions. Using this method reduces the error associated with the limited number of independent $k$-modes that enter the computation of the power in each $k$ bin to only the error on the $\Lambda$CDM power spectrum.

\section{Lensing in coupled dark energy cosmologies}\label{lenscoupled}

Now we present the framework for calculating the gravitational lensing signal in the cDE scenario. The way that light is deflected along the path from its source to an observer is determined by the mass distribution and the geometry of the Universe. The deflections of light lead to distortions of the observed image of the source. The mapping between the original source shape and the observed image is given by
\begin{equation}
	{\cal A}=
	\left(\begin{array}{cc}
	1-\kappa-\gamma_1 & -\gamma_2 \\
	-\gamma_2 & 1-\kappa+\gamma_1
	\end{array}\right),
\end{equation}
\citep[][]{Bartelmann:1999yn} where the convergence, $\kappa$, causes an isotropic dilation and the shear, $\gamma=\gamma_1+i\gamma_2$, changes the ellipticity. $\kappa$ is challenging to measure, as the original size of the source is unknown; equally $\gamma$ cannot be measured for a single source as the intrinsic ellipticity of the source is unknown. However if the shear of a large number of sources is correlated, then the lensing signal can be measured as a correlation function (insofar as the intrinsic ellipticities are not themselves correlated; see discussion in section \ref{Results} below). Therefore we will be interested in the shear correlation function $C_\gamma$ in order to quantify our predictions, given by \citep[][]{Bartelmann:1999yn}
\begin{equation}
	C_\gamma(\theta) = \int ^\infty _0 dl\frac{l}{2\pi}P_\kappa (l)J_0(l\theta)
	\label{eq:ckappa},
\end{equation}
where $\theta$ is the angular distance between the correlated sources, $l$ is the angular wavenumber and the lensing power spectrum $P_\kappa$ is given by \citep[][]{Bacon:2004ht,Massey:2007gh}
\begin{equation}
\label{eq:Pkappa}
	P_\kappa (l) = \frac{9}{4} \left(\frac{H_0}{c}\right)^4 \int ^{\chi _{\rm H}} _0 d \chi W_1(\chi) W_2(\chi) a^4 \Omega_{\rm m}(a)^2 P_\delta \left(\frac{l}{\chi},\chi\right), 
\end{equation}
with the weight functions
\begin{equation}
	W(\chi)=\int^{\chi _{\rm H}} _\chi d\chi' G(\chi')\left(1-\frac{\chi}{\chi'}\right),
	\label{eq:w}
\end{equation}
where $\chi$ is comoving distance, $\chi _{\rm H}$ is the comoving distance to the horizon and $G(\chi)$ is the normalised distribution of the sources in comoving distance, corresponding to a redshift distribution for the sources. We use two weight functions in Equation \ref{eq:Pkappa} since we are using tomographic weak lensing. Equation (\ref{eq:w}) is valid for flat cosmologies, which are all that are considered in this paper. Usually Eq.~(\ref{eq:Pkappa}) is written with the assumption $\Omega_{\rm m}(a)=\Omega_{\rm m}/a^3$; however the form above does not include such an assumption, as coupling CDM and DE means that $\Omega_{\rm m}$ has a different dependence on time, as shown in Eqs.~(\ref{klein_gordon}-\ref{continuity_radiation}).

We have modified the COSMOS CosmoMC code \citep[][]{Lesgourgues:2007te,Lewis:2002ah,Massey:2007gh}, which calculates the combined shear correlation function from the theoretical power spectrum prediction given by CosmoMC, to include cross-correlation of redshift bins and to calculate the predicted weak lensing signal directly from the cDE model power spectra, according to Eqs.~(\ref{eq:ckappa}-\ref{eq:w}). We will now use these results to estimate the discriminatory power from lensing between different coupled DE models. 

\section{Simulations}\label{Simulations}

For our analysis we will rely on the public nonlinear power spectrum data computed from the {\small CoDECS} simulations \citep[][]{CoDECS},
the largest suite of cosmological N-body simulations for cDE models to date, carried out with the modified version by
\citet{Baldi_etal_2010} of the widely used Tree-PM parallel N-body code {\small GADGET} \citep[][]{gadget-2}.
In particular we will consider the {\small H-CoDECS} suite
that includes hydrodynamical simulations of all the cDE models summarized in Table~\ref{tab:models} on relatively small scales.
More specifically, the {\small H-CoDECS} runs follow the evolution of $512^{3}$ CDM and $512^{3}$ gas particles in a cosmological comoving box of $80$ Mpc$/h$
a side, with a mass resolution at $z=0$ of $m_{c}=2.39\times 10^{8}$ M$_{\odot }/h$ for CDM and $m_{b}=4.78\times 10^{7}$ M$_{\odot }/h$ for
baryons, and a force resolution set by the gravitational softening $\epsilon _{g} = 3.5$ kpc$/h$.
Adiabatic hydrodynamical forces on the gas particles are computed by means of the entropy conserving formulation of
{\em Smoothed Particle Hydrodynamics} \citep[SPH,][]{Springel_Hernquist_2002} and other radiative processes such as
gas cooling, star formation, or feedback mechanisms are not included in the simulations.

Initial conditions are generated at $z_{i} = 99$ by rescaling, with the appropriate growth factor for each specific model,
the displacements obtained for a particular random field realization of the linear power spectrum $P_{\rm lin}(k)$ at $z_{\rm CMB}$. This power spectrum is
computed by the publicly available Boltzmann code {\small CAMB} \citep[][]{camb} for a $\Lambda $CDM cosmology
with parameters consistent with the latest ``CMB only Maximum Likelihood" constraints from WMAP7 \citep[][]{wmap7},
which are summarized in Table~\ref{tab:parameters}.
This means that all the different simulations have exactly the same initial conditions at $z_{\rm CMB}$, and their different features at low redshifts
depend uniquely on the different cosmology in place between last scattering and the present time.
\begin{table}
\begin{center}
\begin{tabular}{cc}
\hline
Parameter & Value\\
\hline
$H_{0}$ & 70.3 km s$^{-1}$ Mpc$^{-1}$\\
$\Omega _{\rm CDM} $ & 0.226 \\
$\Omega _{\rm DE} $ & 0.729 \\
${\cal A}_{s}(\sigma_8)$ & $2.42 \times 10^{-9}$ (0.801 for $\Lambda$CDM)\\
$ \Omega _{b} $ & 0.0451 \\
$n_{s}$ & 0.966\\
\hline
\end{tabular}
\end{center}
\caption{The set of cosmological parameters at $z=0$ assumed for all the models included in the {\small CoDECS} project, consistent with the latest results of the WMAP collaboration for CMB data alone \citep[][]{wmap7}.}
\label{tab:parameters}
\end{table}

The {\small H-CoDECS} matter power spectra have been computed by evaluating the density of the different matter components on a grid
with the same size of the PM grid (\ie $512^{3}$ grid nodes) through a Cloud-in-Cell mass assignment of the different matter species and of the total matter distribution. This procedure
allows us to compute the power spectrum up to scales corresponding to the Nyquist frequency of the grid, \ie
$k_{\rm Ny} = \pi N/L \approx 20.0\, h/$Mpc. Beyond this limiting frequency, the power spectrum has been computed with the folding method of
\citet{Jenkins_etal_1998,Colombi_etal_2008}, and the two estimations have been smoothly interpolated around $k_{\rm Ny}$.
Finally, the combined power spectrum has been truncated at scales where the shot-noise reaches $10\%$ of the measured power.

With the power spectra computed with the procedure just described, we have investigated how future weak lensing probes could
perform in constraining cDE cosmologies, as discussed in the next Section.

\section{Results}\label{Results}

\begin{table}
\centering
\begin{tabular}{cccc}
\hline
	 & $n$/ & Area/ \\
  Survey & galaxy arcmin$^{-2}$ & degree$^2$\\
\hline
  DES & 13 & 5000 \\
  Euclid & 30 & 15000 \\
\hline
\end{tabular}
\caption{Galaxy density, $n$, and area assumed for our fiducial DES and Euclid surveys.}
\label{table:galDensity}
\end{table}

We calculated the combined shear correlation function for each of our models using equations \ref{eq:ckappa}-\ref{eq:w}. We consider two types of survey: a ground-based survey modelled on DES, and a space-based survey, Euclid; the adopted galaxy density and survey area are shown in Table \ref{table:galDensity}. In calculating the shear correlation function for these surveys we therefore use a DES-like redshift distribution given by
\begin{equation}
n(z)=(z^a+z^{ab})/(z^b+c)
\end{equation}
where $a=0.612$, $b=8.125$, $c=0.62$, and a space survey redshift distribution for Euclid given by
\begin{equation}
n(z)=\alpha\Sigma_0\frac{z^2}{z_0^3}\exp(-(z/z_0)^\beta)
\end{equation}
where $\alpha=2$, $\beta=3/2$, $z_0=0.63$ and $\Sigma_0=27$ as used in  \citet{Beynon:2009yd}. We also calculated simulated covariance matrices including sample variance and shape noise in a similar way to that calculated in \citet{Beynon:2009yd} using the Horizon simulation \citep[][]{Teyssier:2009zd}; here we used 81 patches of 3.4 square degrees to estimate cosmological sample variance, and assumed an intrinsic shape noise of $\sigma_\gamma=0.2$ on each component of the shear.

In order to examine whether interacting dark energy models can be detected by forthcoming space and ground-based missions, we can assess the difference in $\chi^2$ between the best-fit $\Lambda$CDM and best-fit interacting DE model for a given dataset.  One could choose a fiducial $\Lambda$CDM shear correlation function with realistic error-bars, and find the best-fit interacting DE model for this; but it is more convenient computationally to choose a fiducial interacting DE model and vary parameters of the easily obtained $\Lambda$CDM models to find the best standard cosmology fit. The difference in $\chi^2$ between the two best-fit models is the same whichever way round we choose, and is a measure of our ability to distinguish between the two types of model. 

We ran CosmoMC to find the best fit $\Lambda$CDM models for each of the cDE models with different CDM couplings. We used the following parameter space: $0\leq\Omega_m\leq0.5$, $0.5\leq\sigma_8\leq1$, $0.4\leq h\leq1$, $-2\leq w \leq0$ and $0.01\leq\Omega_b\leq0.15$. The tomographic lensing results were studied for 3 cross-correlated redshift bins of equal size between $z=0.3$ and $z=1.5$ and $1' \leq \theta \leq 90'$.

\subsection{Constant coupling models with an exponential potential}

\begin{figure*}
\centering
\hspace{3mm}
\mbox{
{\psfig{figure=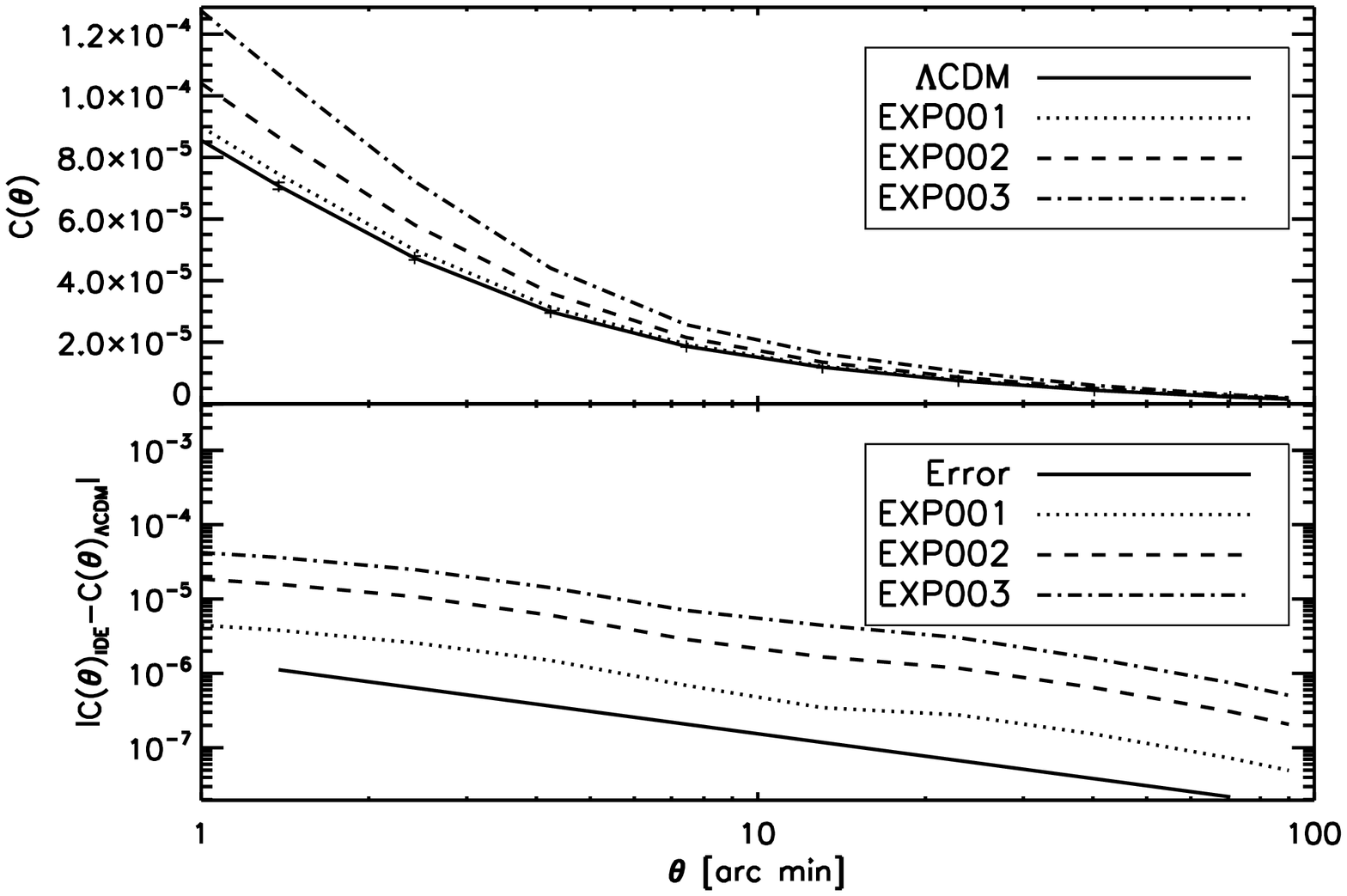,width=76mm}}
}
\hspace{7 mm}
\mbox{
{\psfig{figure=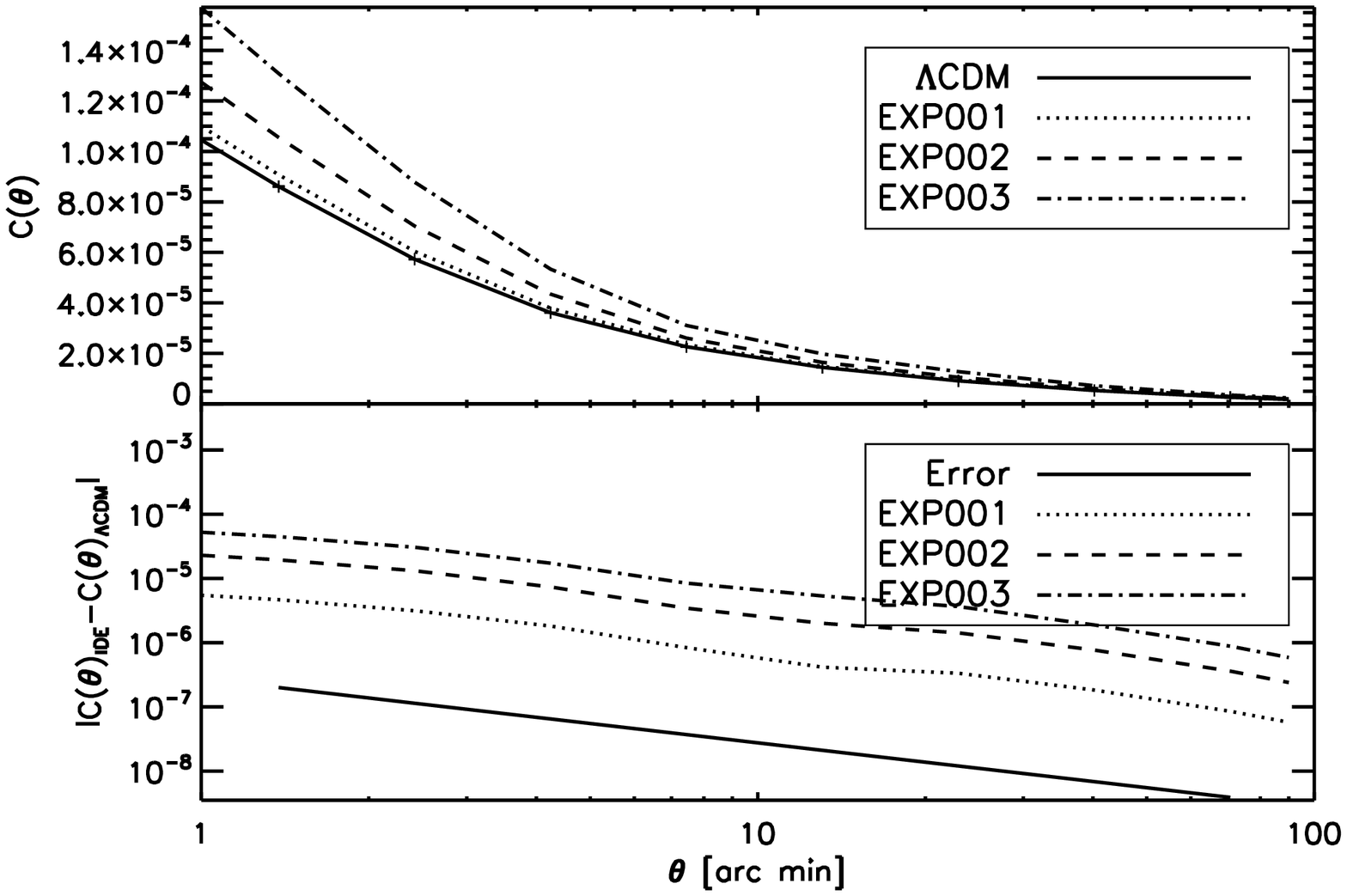,width=76mm}}
}
 \caption{Correlation function predicted for cDE models with error estimates for DES (left) and Euclid (right) surveys using WMAP7 best fit parameters.}
 \label{fig:correlation}
\end{figure*}

\begin{table}
\centering
\begin{tabular}{cccc}
\hline
	 & & DES & Euclid \\
	 Model & $\beta_0$ & $\Delta\chi^2$ & $\Delta\chi^2$\\
\hline
         EXP001&0.05 & 3 & 30 \\
         EXP002&0.1 & 48 & 480\\
	EXP003&0.15 & 340 & 3300 \\
\hline
\end{tabular}
\caption{Best fit $\Delta\chi^2$ for different couplings, using errors calculated for DES and Euclid surveys.}
\label{chi sq table}
\end{table}

\begin{figure*}
\centering
\mbox{
\subfigure[$\Lambda$CDM]{\psfig{figure=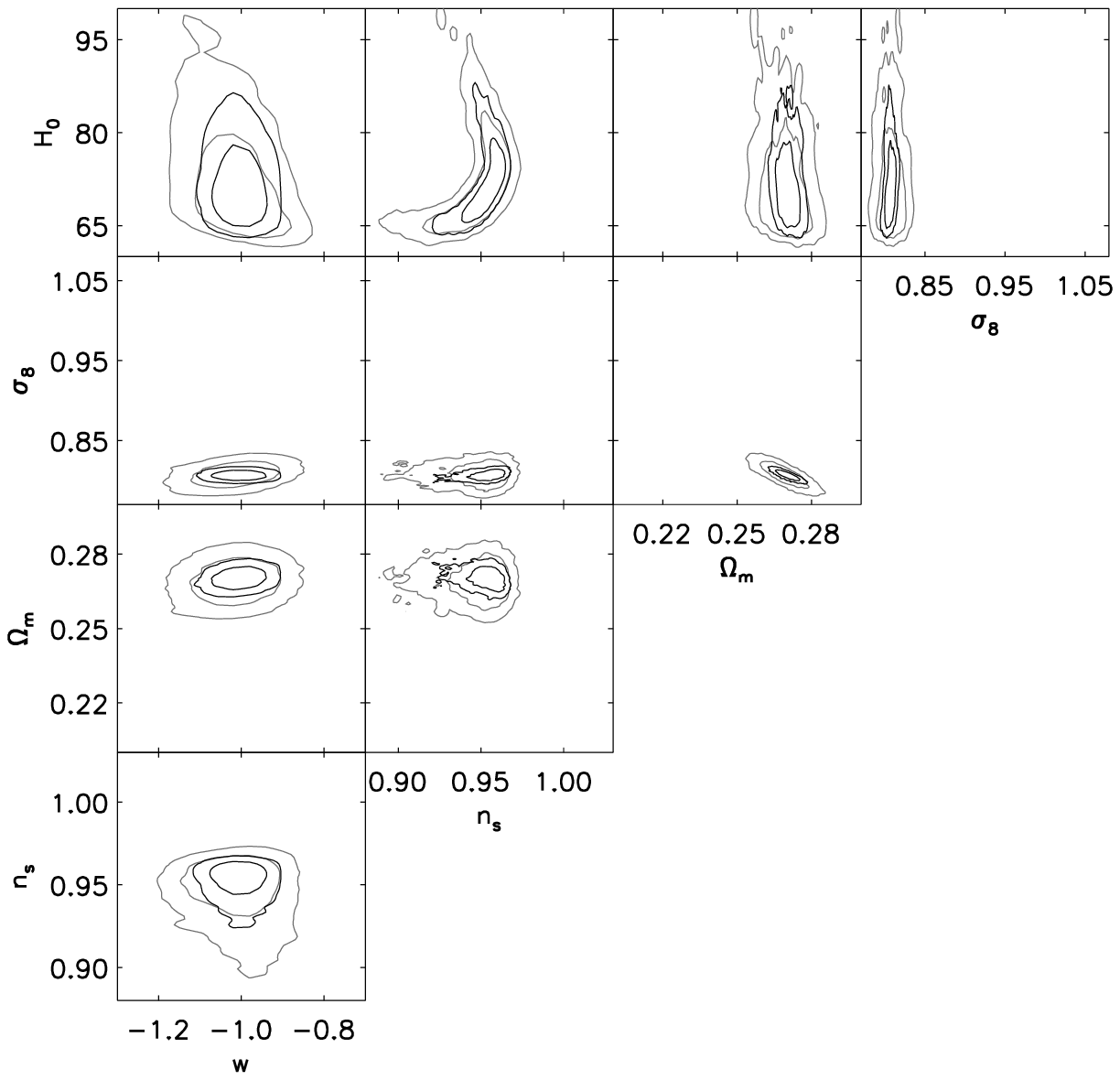,width=110mm} \label{fig:lcdm}} \hspace{-22 mm}
\subfigure[EXP001 ($\beta_0=0.05$)]{\psfig{figure=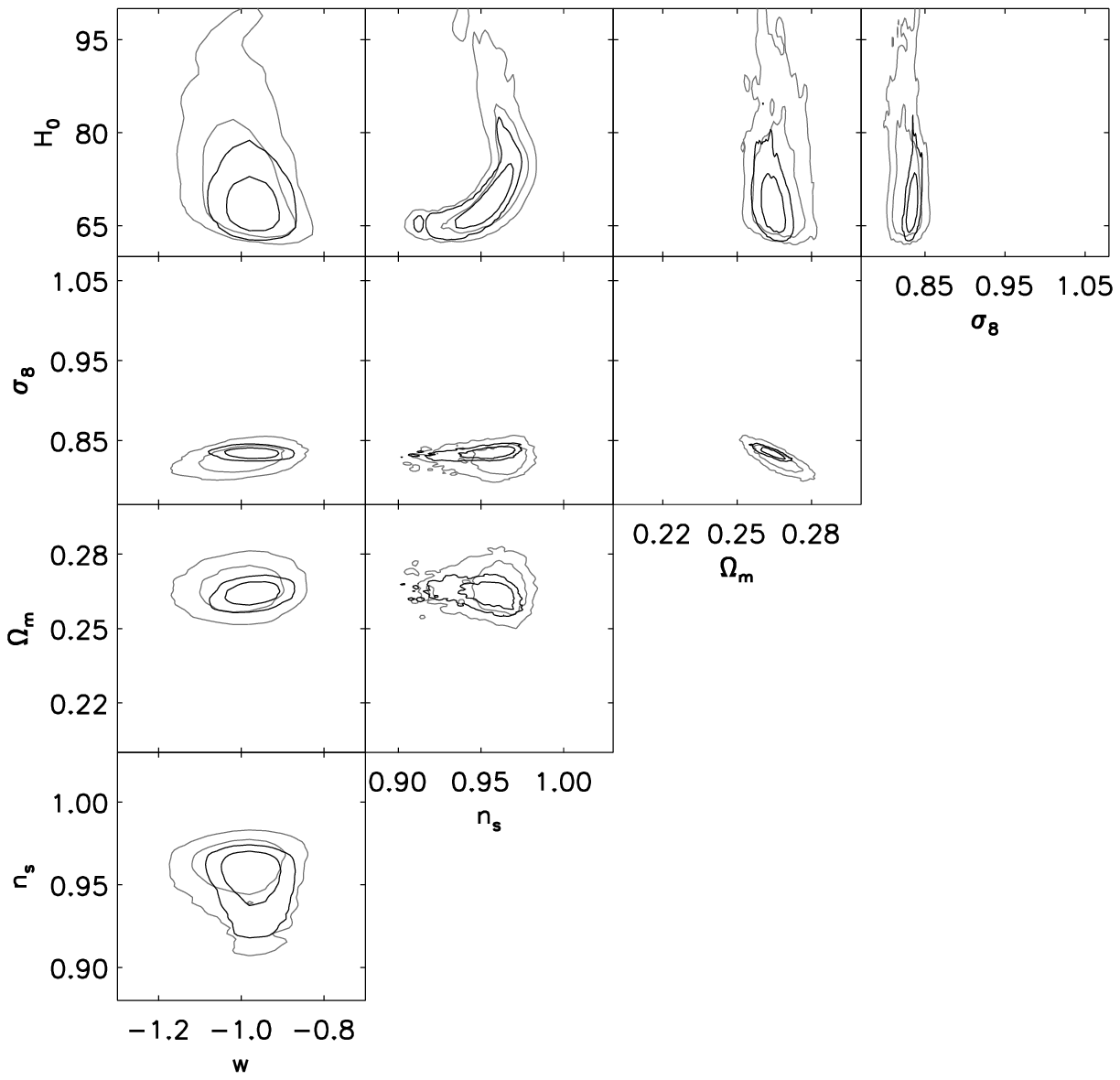,width=110mm} \label{fig:exp001}}
}
\mbox{
\subfigure[EXP002 ($\beta_0=0.1$)]{\psfig{figure=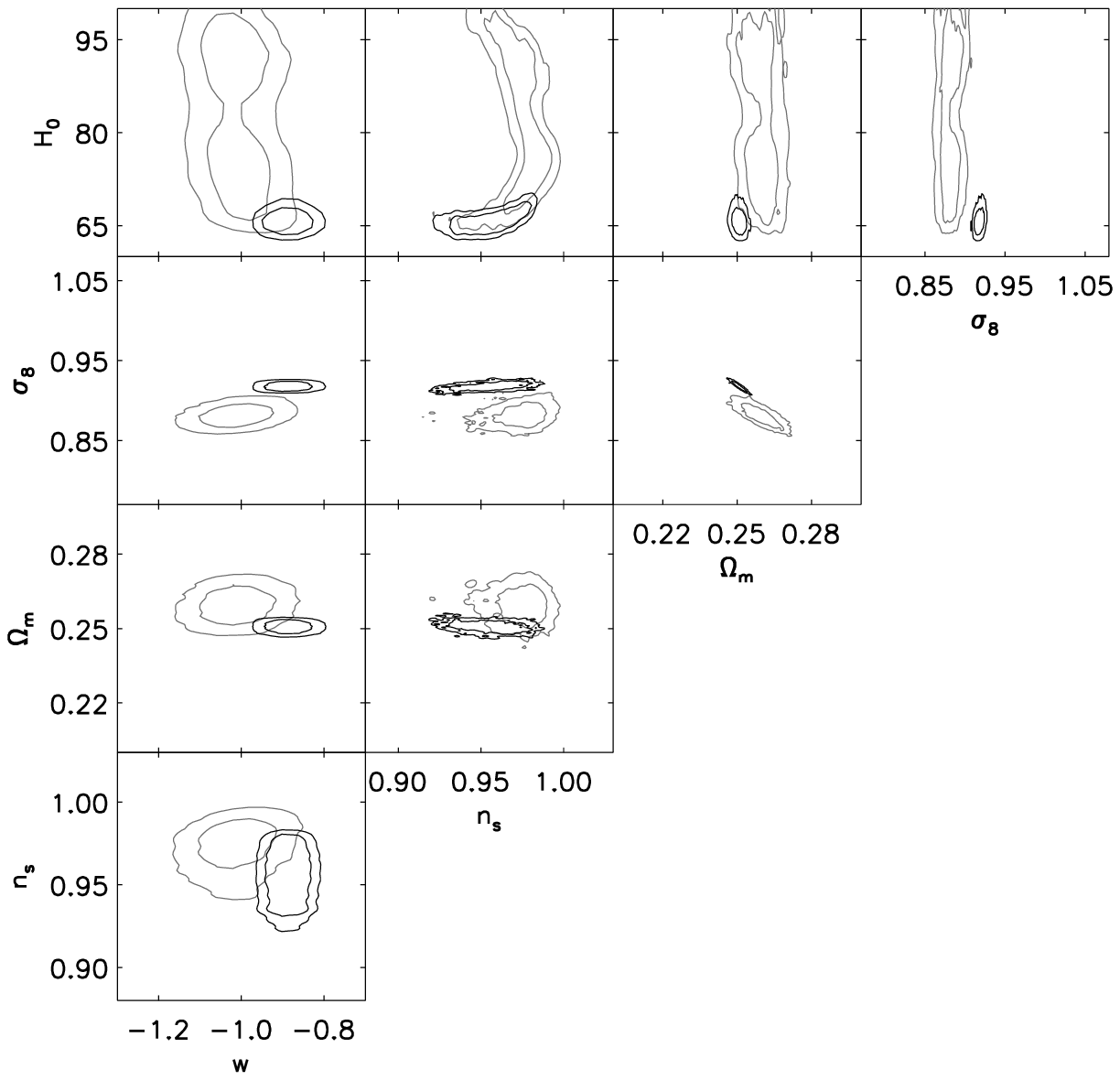,width=110mm} \label{fig:exp002}} \hspace{-22 mm}
\subfigure[EXP003 ($\beta_0=0.15$)]{\psfig{figure=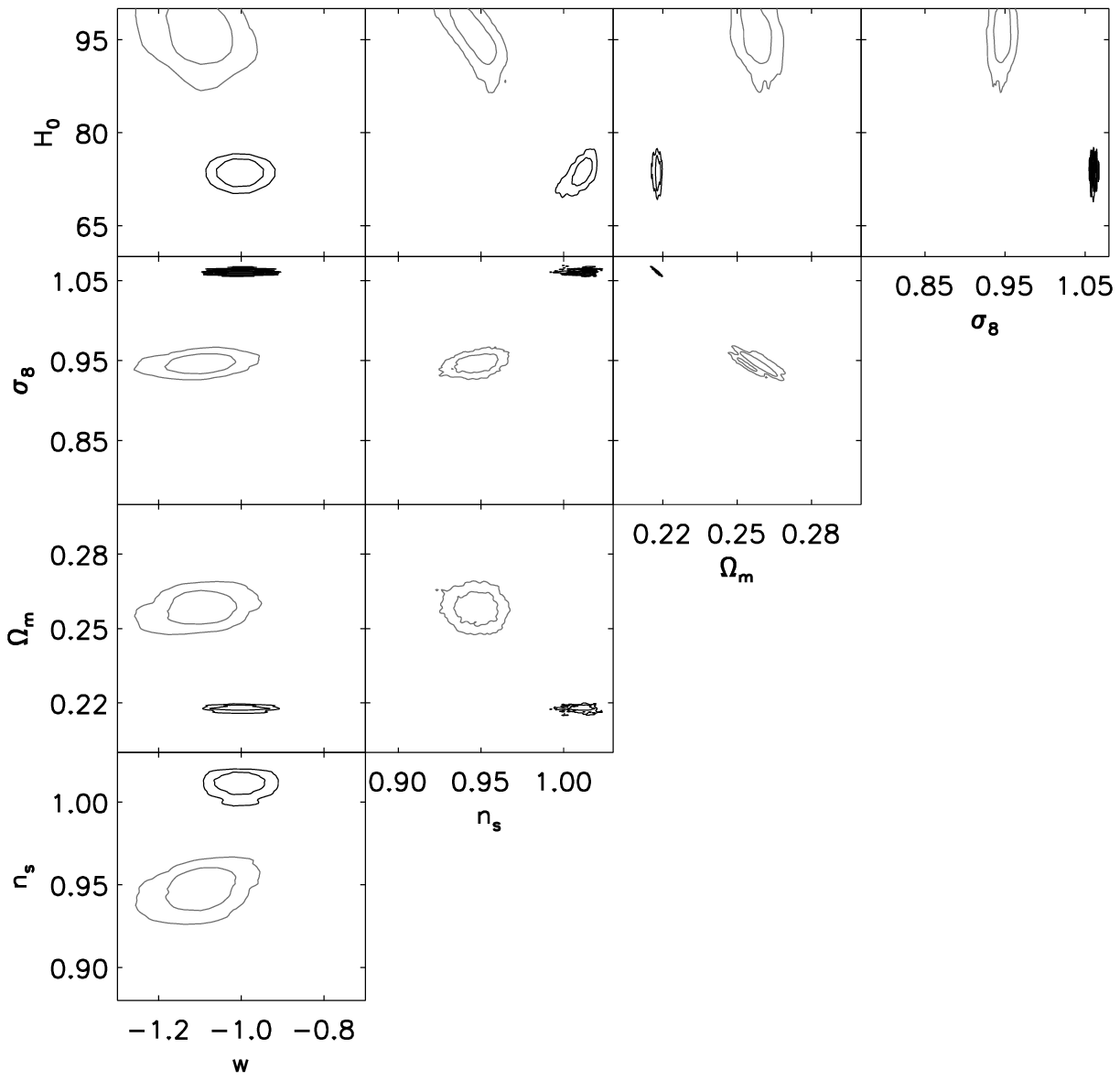,width=110mm} \label{fig:exp003}}
}

\caption{Constraints on $\Omega_m$, $\sigma_8$, $n_s$, $w$ and $H_0$. The light grey contours show the 68\% and 95\% confidence limits for DES, while the dark grey contours show the 68\% and 95\% confidence limits for Euclid.}
\label{fig:contours}
\end{figure*}

In this section we look at how introducing a constant coupling between DM and DE (models EXP001-3 in Table \ref{tab:models}) affects the weak lensing signal. The shear correlation functions, with WMAP7 initial conditions, are shown in Figure \ref{fig:correlation}. Note that $\beta_0$ primarily changes the amplitude of the correlation function, with an additional slight alteration in slope. The difference in $\chi^2$ for each of the different constant couplings is shown in Table \ref{chi sq table}, and we see that lensing with Euclid should be able to discriminate between $\beta_0\geq0.05$ and $\Lambda$CDM at a confidence level of $5\sigma$, while DES should be able to discriminate between $\beta_0\geq0.1$ and $\Lambda $CDM at a confidence level of $4\sigma$.

Figure \ref{fig:contours} shows that the best fit $\Lambda$CDM models for each of the couplings occupy quite different parameter regions, especially for Euclid. The discrepancies between DES and Euclid predictions in these plots are found to be due to the off-diagonal covariance matrix terms; this can be seen by examining the best fit models for DES and Euclid along with the cDE model we are trying to fit. The best fit for our DES survey appears to be a worse fit at small $\theta$ and a better fit at large $\theta$ than the Euclid best fit. This is due to the covariance being largest for large angles and high redshifts. So while DES has a larger contribution from shape noise at small $\theta$ allowing a worse fit on small scales, conversely Euclid is more sensitive to the covariance on large scales. This descrepancy between the DES and Euclid best fit $\Lambda$CDM increases as $\beta_0$ increases.

These results show that if dark energy and dark matter truly do interact in the way described by our class of models, and we attempt to fit a $\Lambda$CDM cosmology to the observations, then we will infer increased values of $H_0$ and $\sigma_8$, and a decrease in $w$ and $\Omega_m$ as $\beta_0$ increases.

It should be noted that \cite{Kirk:2011sw} and \cite{Laszlo:2011sv} have recently shown that the effects of modified gravity and alternative dark energy models can be degenerate with systematics due to intrinsic alignments. Baryonic physics has also been shown to have possibly large effects on the matter power spectrum from scales as small as $k$=0.3 h/Mpc \citep[][]{vanDaalen:2011xb,Semboloni_etal_2011}. In this paper we do not include these effects, as we are seeking to present the pure shear signal predictions. Our results should therefore be considered best-case predictions which will be diluted by the impact of systematic and baryonic effects.

\begin{table*}
\begin{tabular}{llccccc}
\hline
Survey & Model & $w$  & $H_0 $ & $\sigma_8$ & $\Omega_m$ & $n_s$ \\
\hline
\multirow{5}{*}{DES} &EXP001 & $-0.974 \pm 0.020$ & $69.2\pm3.5$ & $0.834\pm0.005$ & $0.264\pm0.003$ & $0.952\pm0.013$\\
& EXP002 & $-1.012\pm0.047$ & $82.7\pm9.9$ & $0.881\pm0.010$ & $0.259\pm0.005$ & $0.973\pm0.012$\\
& EXP003 & $-1.110\pm0.045$ & $95.1\pm2.8$ & $0.946\pm0.008$ & $0.258\pm0.004$ & $0.947\pm0.009$ \\
& EXP008e3 & $-0.981\pm0.048$ & $77.3\pm10.0$ & $0.889\pm0.010$ & $0.262\pm0.005$ & $0.954\pm0.014$ \\
& SUGRA003 & $-0.755\pm0.044$ & $81.1\pm6.1$ & $0.760\pm0.013$ & $0.305\pm0.008$ & $0.760\pm0.013$ \\
\hline
\multirow{5}{*}{Euclid} &EXP001 & $-0.974\pm0.020$ & $69.2\pm3.5$ & $0.834\pm0.005$ & $0.264\pm0.003$ & $0.952\pm0.013$ \\
& EXP002 & $-0.888\pm0.020$ & $66.1\pm1.5$ & $0.918\pm0.004$ & $0.251\pm0.002$ & $0.956\pm0.018$ \\
& EXP003 & $-1.004\pm0.020$ & $73.3\pm1.3$ & $1.060\pm0.002$ & $0.218\pm0.001$ & $1.009\pm0.007$ \\
& EXP008e3 & $-0.881\pm0.020$ & $65.6\pm0.5$ & $0.935\pm0.004$ & $0.247\pm0.002$ & $0.922\pm0.016$ \\
& SUGRA003 & $-0.804\pm0.020$ & $85.4\pm2.2	$ & $0.745\pm0.004$ & $0.314\pm0.004$ & $1.092\pm0.007$ \\
\hline
\end{tabular}
\caption{Marginalised parameters for $\Lambda$CDM fit to models for DES and Euclid surveys with $1\sigma$ errors.}
\label{tab:margparams}
\end{table*}

\subsection{Other potentials and coupling}

\begin{figure}
\hspace{7mm}
\psfig{figure=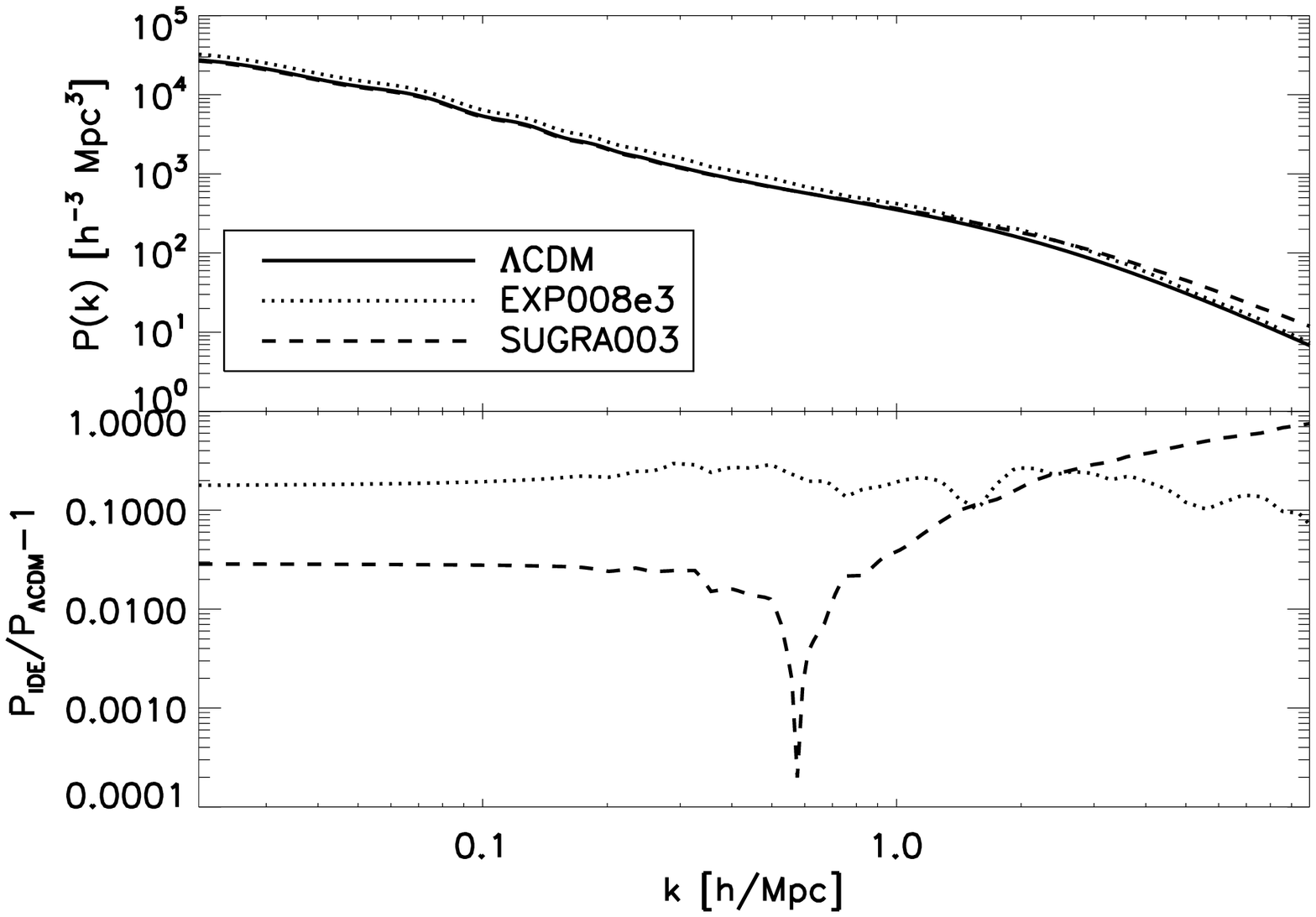,width=76mm}
 \caption{Power spectrum for an evolving coupling model with an exponential potential (EXP008e3), and a constant coupling model with a SUGRA potential.}
 \label{fig:PowerspectrumEvolve}
\end{figure}

Although for the previous section we restricted ourselves to looking at constant coupling models with an exponential potential, the cDE model has the freedom to examine different potentials and an evolving coupling. Two of the {\small CoDECS} simulations explore this freedom: EXP008e3, which has the same potential as the models in the previous section but with an evolving coupling, and SUGRA003, which has a SUGRA potential with a constant coupling. Since there is not yet a suite of these types of simulations exploring the full range of parameter space, we have included them as lone examples simply to demonstrate the range of the cDE model. The power spectrum for these models is shown in Figure \ref{fig:PowerspectrumEvolve}, where we can see that for the EXP008e3 model we get similar differences between the cDE model and $\Lambda$CDM to those shown in the larger constant coupling models (EXP002/3). On the other hand, the SUGRA003 model has smaller differences to this at large scales and much larger differences at small scales (almost 100\% at $k=10h/Mpc$) demonstrating how important it is to carry out full simulations of these models in order to obtain small scale predictions.

We again attempted to find a best fit $\Lambda$CDM model using CosmoMC and the $\chi^2$ for the best fit result, shown in Table \ref{chi sq table other}, demonstrates that for these particular models we would be able to exclude both models at $>7\sigma$ for both DES and Euclid. Further investigation of these types of model would allow constraints to be made on the parameters characterizing the coupling and the potential.

\begin{table}
\centering
\begin{tabular}{ccc}
\hline
	 & DES & Euclid \\
	 Model & $\Delta\chi^2$ & $\Delta\chi^2$\\
\hline
         EXP008e3& 64 & 570 \\
         SUGRA003& 16 & 100 \\
\hline
\end{tabular}
\caption{Best fit $\Delta\chi^2$ for EXP008e3 and SUGRA003 using errors calculated for DES and Euclid.}
\label{chi sq table other}
\end{table}

\subsection{Comparison of simulations and Halofit}
\label{comparison}

\begin{figure*}
\centering
\hspace{3mm}
\mbox{
{\psfig{figure=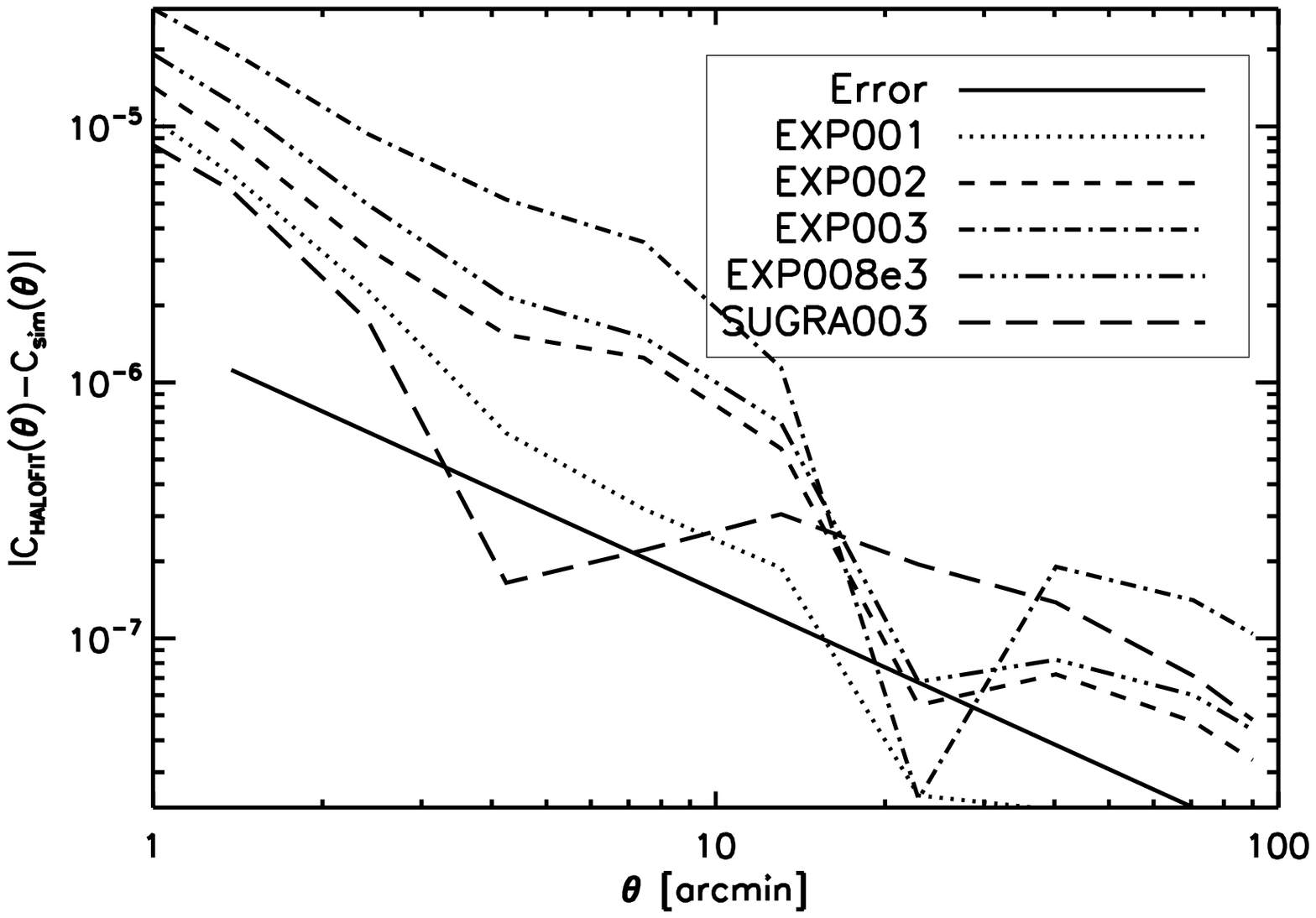,width=76mm}}
}
\hspace{7 mm}
\mbox{
{\psfig{figure=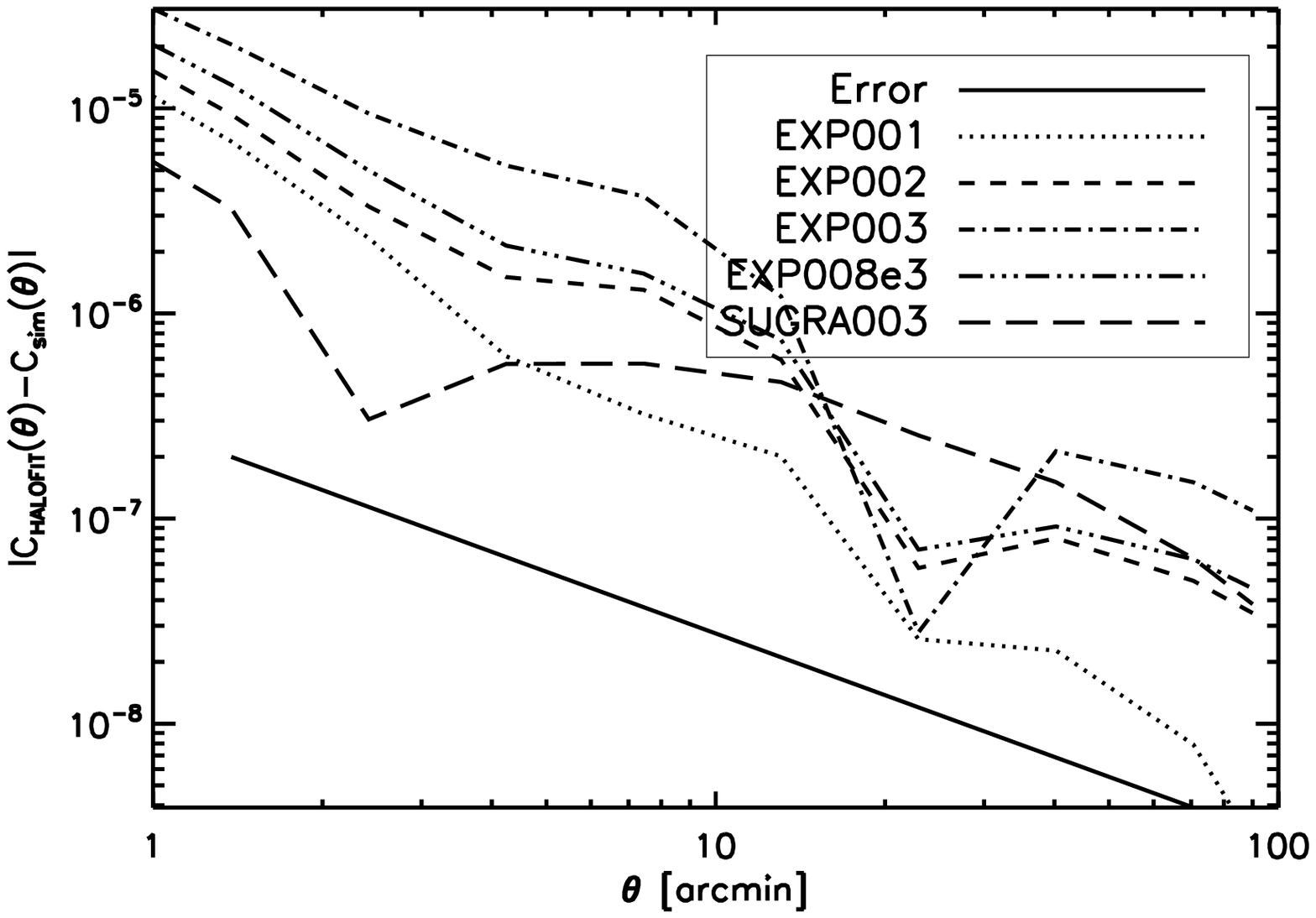,width=76mm}}
}
 \caption{Difference between shear correlation function calculated using simulations and shear correlation function calculated using Halofit. Also shown is the measurement error (from sample variance and shape noise) for DES (left) and Euclid (right) using WMAP7 best fit parameters.}
 \label{fig:correlation_halo}
\end{figure*}

In section \ref{coupledDE} we discussed the importance of using N-body simulations over using $\Lambda$CDM non-linear fitting formulas such as Halofit \citep[][]{Smith:2002dz} to estimate the non-linear power spectrum for cDE models. In Figure \ref{fig:correlation_halo} we show that the use of Halofit to estimate the non-linear power spectrum results in errors in the shear correlation function that exceed the statistical errors, for each of the surveys and for all of the models considered. This demonstrates the importance of using N-body simulations to predict the non-linear matter power spectrum for cDE models, and that further simulations for a variety of cDE models are required to make accurate weak lensing forecasts using non-linear scales. 

\section{Conclusions}\label{Conclusions}

In this paper we have presented weak lensing predictions for cDE models using the non-linear power spectrum calculated by the {\small CoDECS} simulations.

We have calculated the total shear power spectrum for each of the models, and used CosmoMC to find  the best fit $\Lambda$CDM model; we have demonstrated the discriminatory power of future lensing surveys such as DES and Euclid, where it should be possible to tightly constrain constant coupling models with exponential potentials to $\beta_0<0.05$ with Euclid, or $\beta_0<0.1$ with DES. However, this should be considered a best-case scenario, since the inclusion of intrinsic alignments and baryonic physics may impact the constraining power; this will be the subject of future work.

We have shown that for cDE models with larger coupling there is a clear difference between the best fit $\Lambda$CDM for the same model but different surveys. This difference is due to the dominance of the off-diagonal covariance matrix terms over the diagonal for larger surveys, and shows the importance of including these off-diagonal terms in weak lensing predictions.

We have also calculated the expected signal for a non-constant coupling model and a non-exponential potential model. These models could be excluded by $\geq2\sigma$ for a DES-like survey and $>7\sigma$ for Euclid. However we have not obtained constraints on the parameters of these types of model, since currently N-body simulations for these models have only been run with one parameter set. A substantial set of simulations would be required in order to properly sample the parameter space of these more complex scenarios. This will be a worthwhile task, as the effects of these cosmologies appear to be more difficult to detect in the background and in the linear regime with respect to standard interacting dark energy models, making non-linear N-body simulations vital for realistic lensing predictions.

We have also shown the size of the error on weak lensing predictions if a $\Lambda$CDM non-linear fitting formula, such as Halofit, is used to estimate the matter power spectrum, instead of using simulations. We find that this Halofit error is larger than the statistical error for the DES and Euclid surveys, and for all the models considered here. This demonstrates the importance of using a full N-body code to estimate the non-linear power spectrum.

\section*{Acknowledgements}
DB is supported by an RCUK Academic Fellowship. KK is supported by the STFC (grant no. ST/H002774/1), a European Research Council Starting Grant 
and the Leverhulme trust. EB is funded by an STFC PhD studentship. MB acknowledges support by 
the DFG Cluster of Excellence ``Origin and Structure of the Universe'' and by the TRR33 Transregio Collaborative Research Network on the ``Dark Universe''.

\label{lastpage}

\end{document}